\documentclass[prl,aps,twocolumn]{revtex4}

\usepackage{graphicx}

\begin{document}

\title{Topological defects in flat nanomagnets: the magnetostatic limit}

\author{G.-W. Chern}

\author{H. Youk}

\author{O. Tchernyshyov}

\affiliation{Department of Physics and Astronomy, 
The Johns Hopkins University, 
3400 N. Charles St., 
Baltimore, Maryland 21218}

\begin{abstract}
We discuss elementary topological defects in soft magnetic
nanoparticles in the thin-film geometry.  In the limit dominated by
magnetostatic forces the low-energy defects are vortices (winding
number $n = +1$), cross ties ($n = -1$), and edge defects with $n =
-1/2$.  We obtain topological constraints on the possible composition
of domain walls.  The simplest domain wall in this regime is composed
of two $-1/2$ edge defects and a vortex, in accordance with
observations and numerics.
\end{abstract}

\maketitle

Nanorings made out of a soft ferromagnetic material generate
considerable interest as prospective building blocks for nonvolatile
random-access memory \cite{Zhu00}.  An attractive feature of the ring
geometry is the existence of two lowest-energy states in which
magnetization points in the azimuthal direction (clockwise or
anticlockwise) preventing the straying of magnetic field.  The
switching between the two states can be accomplished by applying the
magnetic field in the plane of the ring or by injecting electric
current.  In both cases the switching is accomplished by nucleating a
small bubble of the opposite domain and letting it expand until it
occupies the entire ring \cite{FQZhu04}.  Alternatively one can view
the process as the creation, propagation, and mutual annihilation of
two {\em domain walls} separating the domains with clockwise and
counterclockwise magnetization.  These considerations motivate us to
study the properties of domain walls in nanorings.

Domain walls in magnetic nanoparticles differ substantially from
domain walls in macroscopically large magnets.  The main reason for
that is the more prominent role of the surface in smaller samples.
Qualitative changes are expected when one (or more) of the particle
dimensions crosses a length scale characterizing the strength of
ferromagnetic exchange relative to that of the stray field, $\lambda =
\sqrt{A/\mu_0 M_0^2}$, or material anisotropy, $\lambda_a =
\sqrt{A/K}$.  Here $A$ is the exchange constant, $M_0$ is the
equilibrium magnetization, and $K$ is the anisotropy constant of the
material \cite{Hubert}.  The anisotropy scale $\lambda_a$ is
particularly large in soft materials and can be considered infinite
for submicron particles.  The exchange length $\lambda$ is in the
range of a few nanometers.

Previous experimental and numerical studies of domain walls in
submicron rings \cite{Klaeui03} demonstrate that the ring curvature
does not have a significant impact on the properties of domain walls.
We therefore discuss the simpler geometry of a strip.  Domain walls in
strips were studied numerically by McMichael and Donahue
\cite{McMichael97} who found (at least) two different types:
``transverse walls'' in extremely thin and narrow strips and ``vortex
walls'' in thicker and wider ones.  Two of us \cite{OT05} have
previously shown that the transverse walls are {\em composite} objects
made of two elementary topological defects located at the opposite
edges of the strip.  In the limit of an extremely thin and
(reasonably) narrow strip the energy of a domain wall comes mostly
from the exchange interaction, so that the magnet is described by the
two-dimensional XY model with an anisotropy at the edge
\cite{Kurzke04}.  In this limit, the elementary topological defects
are (a) vortices and antivortices in the bulk of the strip carrying
winding numbers $+1$ and $-1$, respectively, and (b) halfvortices, or
``boundary vortices'' \cite{Kurzke04}, confined to the edge and
carrying {\em fractional} winding numbers $\pm 1/2$.  A transverse
domain wall consists of two halfvortices with opposite winding numbers
\cite{OT05}.

In this paper and its companion \cite{MMM2} we discuss the structure
and energetics of domain walls in the limit dominated by the
magnetostatic interaction, achieved in strips whose width and
thickness substantially exceeds the exhange length $\lambda$.  We
demonstrate that the ``vortex walls'' observed in this limit are also
composite objects containing {\em three} elementary topological
defects: a vortex (winding number $+1$) residing between two edge
defects (winding numbers $-1/2$).  The identification of the
elementary topological defects and implications for the composition of
domain walls is the subject of this paper.  The companion paper
\cite{MMM2} deals with the energetics of composite domain walls.

Vortex walls are stabilized when both the width and thickness of a
strip substantially exceed the exchange length $\lambda$
\cite{McMichael97}.  In this limit the magnetostatic energy is the
dominant contribution to the energy of a domain wall
\cite{Kohn04,DeSimone-preprint} and the primary force determining the
shape of topological defects.  Because the magnetostatic energy is a
nonlocal functional of magnetization \cite{Hubert}, energy
minimization is a computationaly difficult problem.  Therefore
identification of topological defects is not as straightforward as in
the limit dominated by exchange \cite{OT05}.  Furthermore, the
magnetostatic energy has a large number of absolute minima and one
must search among these solutions for one with the lowest exchange
energy, making this a degenerate perturbation problem.

For simplicity we will use the geometry of a thin film with a constant
thickness $t$ that is small in comparison to the width of the strip
$w$.  In this case the shape anisotropy forces the magnetization
$\mathbf M$ to lie in the plane of the film (with the possible
exception of vortex cores \cite{Wachowiak02}).  It will be further
assumed that the magnetization depends on the coordinates in the plane
of the film only,
\begin{equation}
\mathbf{M} = \mathbf{M}(x,y) = (M_0 \cos{\theta}, M_0 \sin{\theta}, 0).
\label{eq-planar}
\end{equation}

For a given configuration of magnetization ${\bf M}({\bf r})$ its
magnetostatic energy $(\mu_0/2) \int H^2 \, dV$ can be recast as the
Coulomb energy of magnetic charges with density $\rho_m({\bf r}) =
-\nabla \cdot \mathbf{M} = M_0(\sin{\theta} \, \partial_x \theta -
\cos{\theta} \, \partial_y \theta)$.  Being positive definite, the
magnetostatic energy has an absolute minimum of zero, which
corresponds to the complete absence of magnetic charges.  Thus it
makes sense to look for low-energy states with topological defects
among configurations with zero charge density in the bulk, $-\nabla
\cdot \mathbf{M} = 0$, and on the surfaces, $\hat\mathbf{n} \cdot
\mathbf{M} = 0$ (here $\hat\mathbf{n}$ is the surface normal).  A
method for constructing such solutions has been discussed by van den
Berg \cite{Berg86}.  It yields configurations with domains of smoothly
varying magnetization separated by discontinuities in the form of
Neel-type domain walls.  The walls acquire a finite width when the
exchange interaction is taken into account.

\begin{figure}
\includegraphics[width=0.9\columnwidth]{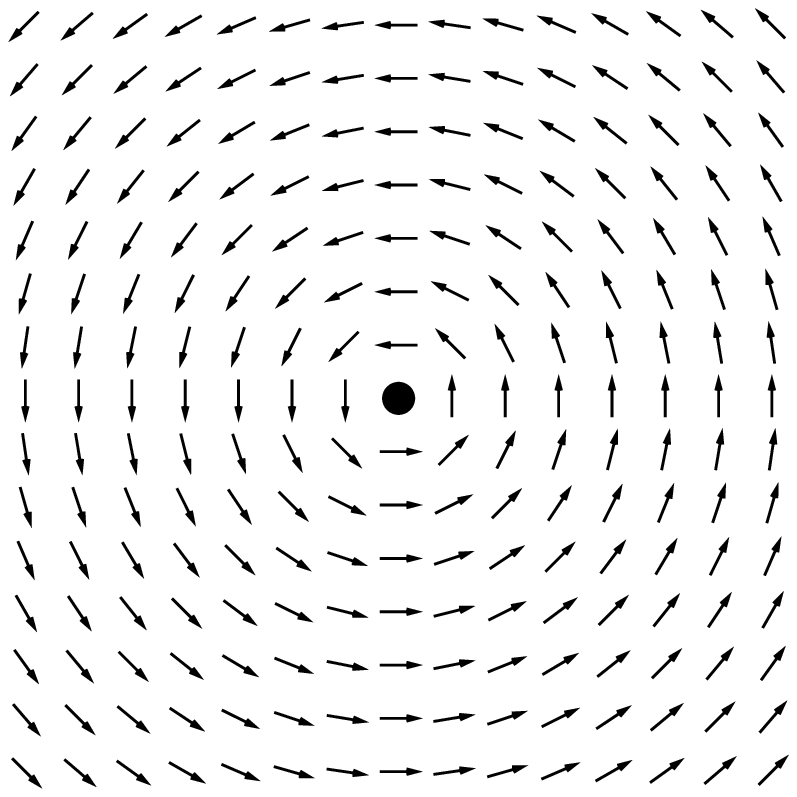}
\includegraphics[width=0.9\columnwidth]{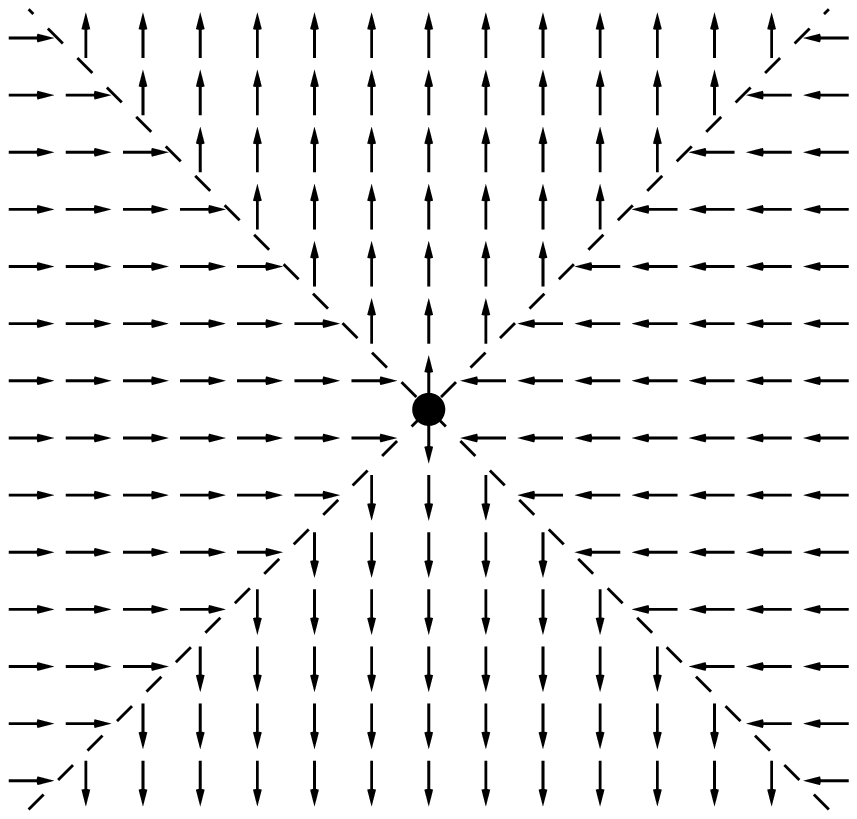}
\includegraphics[width=0.9\columnwidth]{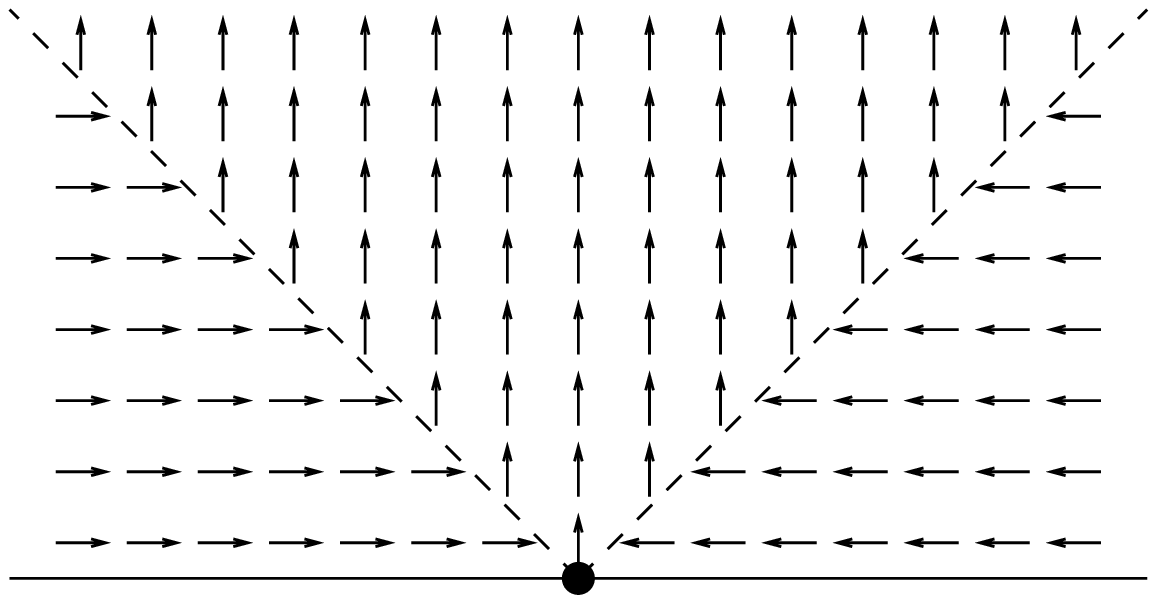}
\caption{Top to bottom: a vortex, an antivortex, and a $-1/2$ edge defect
in the magnetostatic limit.}
\label{fig-anti}
\end{figure}

{\bf Vortex.}  The simplest nontrivial example is the configuration
with a single vortex at the origin, $\exp{(i\theta(x,y))} = \pm
i(x+iy)/|x+iy|$.  The two signs give two different values of {\em
chirality} (direction of circulation) of the vortex.  In both cases
the topological charge, or the {\em winding number} \cite{Chaikin}, is
$+1$.  This solution has zero density of magnetic charge and thus
minimizes the magnetostatic term.  Furthermore, it also represents a
local minimum of the exchange energy.  Taken together, these two
observations show that the vortex is a stable configuration.  Its
energy diverges logarithmically with the system size $R$:
\begin{equation}
E_{+1} \sim 2\pi A t \log{(R/\lambda)} + E_{\rm core}.
\label{eq-vortex}
\end{equation}

{\bf Antivortex.}  In contrast, the antivortex configuration
minimizing exchange energy, $\exp{(i\theta(x,y))} = \pm
i(x-iy)/|x-iy|$, has a nonzero density of magnetic charge and thus
represents a poor starting point for constructing a bulk topological
defect with the winding number $-1$ in this limit.  Minimization of
the magnetostatic energy in this topological sector is achieved in the
configuration known as the cross tie \cite{Loehndorf96}, an
intersection of two $90^\circ$ Neel walls normal to each other
(Fig.~\ref{fig-anti}).  The energy of an antivortex grows
linearly with the length of the Neel walls $L$ emanating from it:
\begin{equation}
E_{-1} \sim \sigma t L + E_{\rm core}
\label{eq-antivortex}
\end{equation}
(the core energy is generally different from that of a vortex).  The
surface tension of the wall $\sigma$ is determined by the competition
of exchange and magnetostatic forces.  When the film thickness
$t$ exceeds the Neel-wall width (of order $\lambda$), the calculation
simplifies: the magnetization depends only on the coordinate
transverse to the wall.  The surface tension is then found by
minimizing the total energy per unit area \cite{Hubert}
\begin{equation}
\sigma = \int dx \, \left[A (d\theta/dx)^2 
+ \mu_0 M_0^2 (\cos{\theta}-\cos{\theta_0})^2/2\right],
\end{equation}
subject to the boundary conditions $\theta(\pm \infty) = \pm\theta_0$,
where $2\theta_0$ is the angle of spin rotation across the wall.  
Minimization of the total energy yields a domain wall of a characteristic
width $\lambda\sqrt{2}/\sin{\theta_0}$ with surface tension
\begin{equation}
\sigma = 2\sqrt{2} \, 
(\sin{\theta_0} - \theta_0 \cos{\theta_0}) \, A/\lambda.
\end{equation}
In thinner films ($t \lesssim \lambda$) the magnetostatic term is
nonlocal and the Neel walls acquire long tails \cite{Hubert}.
% $E_{-1} \approx (2-\pi/2)A \, tL/\lambda$.  

{\bf Edge defects.}  An edge defect \cite{OT05} with the winding
number $-1/2$ can be constructed by placing the core of a cross tie at
the edge of the film, so that the magnetization along the edge is
parallel to the boundary (Fig.~\ref{fig-anti}).  As the core is
circumvented counterclockwise the magnetization rotates clockwise
through $\pi$, in agreement with the definition \cite{OT05}.  The
energy of such a defect is also given by Eq.~(\ref{eq-antivortex}).

We have not been able to find any configuration that would contain a
$+1/2$ edge defect and be free from magnetic charges.  It looks likely
that the $+1/2$ defects carry a finite amount of magnetic charge and
thus have a substantially higher magnetostatic energy than the other
three types of defects described above.  This may indicate that, in
the limit where the magnetostatic energy dominates, a $+1/2$ defect
will decay into a vortex ($n = +1$) and an edge defect ($n = -1/2$).
The $+1/2$ defects are stable in the exchange limit
\cite{OT05,Kurzke04}.

{\bf Defects and composite domain walls.}  The defects discussed in
this paper determine the properties of domain walls in nanomagnetic
strips.  Postponing a detailed discussion to the accompanying paper
\cite{MMM2} here we make two general observations that place important
constraints on the possible composition of a domain wall.

\begin{figure}
\includegraphics[width=\columnwidth]{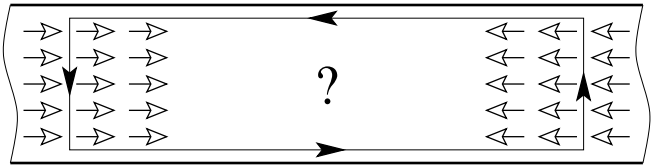}
\caption{Determination of the topological charges at the edges and in
the bulk.}
\label{fig-topology}
\end{figure}

First, a domain wall in a strip must contain (at least) one edge
defect at each edge.  This follows from the definition of their
winding numbers \cite{OT05}.  Moving along the upper/lower edge
(Fig.~\ref{fig-topology}) one finds that the magnetization rotates
through the angle $-2\pi n_{1,2}$.  In the presence of a domain wall,
the edge winding numbers $n_1$ and $n_2$ are half-integers.

Second, the total topological charge of a domain wall, including the
winding numbers of the edges and the bulk, must be zero.  This can be
seen by drawing a contour enclosing the domain wall
(Fig.~\ref{fig-topology}) and noting that the total angle of 
rotation along that contour $-2\pi n_1 - 2\pi n_2$ also equals $2\pi
n$, where $n$ is the winding number in the bulk.  Hence $n + n_1 + n_2
= 0$.

Thus domain walls with the smallest number of defects may contain (a)
two edge defects with winding numbers $+1/2$ and $-1/2$ and no bulk
defects; (b) two $+1/2$ edge defects and one antivortex; and (c) two
$-1/2$ edge defects and one vortex.  Case (a) corresponds to the
transverse wall, which is indeed the lowest-energy domain wall in the
exchange limit \cite{OT05}.  In the opposite magnetostatic limit one
must minimize the number of $+1/2$ edge defects (which have a high
magnetostatic energy).  Therefore it is reasonable to expect that the
lowest-energy domain walls in this limit are of type (c).  Both
experimental observations \cite{Klaeui03} and numerical simulations
\cite{McMichael97} are consistent with this proposition.  See the
companion paper \cite{MMM2} for details.

Much of the recent experimental effort in nanomagnetism has been
devoted to the study of vortices \cite{Shinjo00,Wachowiak02,Choe04}.
Given an equal (if not greater) importance of edge defects in
determining the properties of domain walls, a careful examination of
topological defects at the edge is highly desired.

{\bf Acknowledgment.}  We thank C.-L. Chien, J.-G. Zhu, and F. Q. Zhu
for discussions.  This work was supported in part by the NSF Grant
No. DMR05-20491.

\bibliography{micromagnetics}

\end{document}